\documentclass[11pt]{article}
\usepackage{graphicx}
\usepackage{amsmath}
\usepackage{amssymb}
\usepackage{amsxtra}

\usepackage{array, multirow, hhline}

\def\institute#1{\gdef\@institute{#1}}
\let\oldmaketitle\maketitle
\renewcommand\maketitle{\oldmaketitle\noindent\@institute}

\begin{document}
%%% template for contributors to 'What Comes Beyond the Standard Models'
%%%

\title{Problems of dark atom cosmology}
\author{V. A. Beylin$^{1}$\footnote{vitbeylin@gmail.com}, M. Yu. Khlopov$^{1,2,3}$\footnote{khlopov@apc.univ-paris7.fr}, D. O. Sopin$^{2,3}$\footnote{sopindo@mail.ru}}
\institute{
$^1$ Virtual Institute of Astroparticle physics, 75018 Paris, France\\
$^2$ National Research Nuclear University MEPhI, 115409 Moscow, Russia\\
$^3$  Research Institute of Physics, Southern Federal University, 344090
\\Rostov on Don, Russia\\
}

%%%\begin{document} % for stand-alone document only
\maketitle

\begin{abstract}
The dark atoms $XHe$ are the composite Thomson like atomic dark matter candidates. We address two cosmological problems of this model. The excess of new superheavy particles with even negative charge $X^{-2n}$ over the corresponding antiparticles is balanced by sphaleron transitions with baryon asymmetry and the mass range of $X$ particles should be specified at which this excess can provide dominance of dark atoms in the dark matter density. The other problem is possible capture of light nuclei by dark atoms, which can lead to formation of anomalous isotopes. The possibility of formation of multi dark atom systems at the nucleosynthesis stage is also studied. We approach these open questions of dark atom cosmology in the present work.
\end{abstract}

\section{Introduction}\label{s:intro}

The modern cosmological paradigm involves dark matter (DM). Its existing is confirmed by numerous astrophysical observations: gravitational lensing, anisotropy of the cosmic microwave background and the behavior of galaxies. The theoretical study of cosmological large-scale structures formation also requires to introduce new stable forms of non-relativistic non-baryonic matter.

The nature of DM particles is determined by physics beyond the Standard Model (SM). It was shown that new states could have only even negative electric charge to fulfill the constraints on anomalous isotopes concentration \cite{KhlopovNonGo}. In that scenario the DM density should be provided by the X-helium dark atoms $X^{-2n}(He^{+2})_n$, which forms in two steps. At first of them the excess of heavy negatively charged particles $X^{-2n}$ over the corresponding antiparticles finally generates in sphaleron transitions. The density of the DM could be balanced with the density of baryonic matter \cite{WTCtermo1,WTCtermo2,WTCtermo3}. In particular model the constraints on the mass of the new particles can also be found if it is only one source of the DM. The second stage is the formation of a bound state by capturing the light primordial nuclei. 

However, the description of both steps should be clarified. The properties of the sphaleron transitions are depends on a model significantly \cite{NolteKunz,SpannowskyTamarit}. Therefore calculations with using the SM parameters could be considered only as the first estimation. At the same time, the process of the bound state formation is well described only for small values of charge \cite{Soldatov2010,AkhmedovPospelov}. The serial capture of light nuclei leads to the significant changes in the inner structure of the dark atom. Moreover, such reactions have not been described yet. Also the possible interaction of two bound states at early stages of Universe evaluation has not been studied. In general case, to solve this problems, it is necessary to use the quantum description of Thompson-like dark atoms, which is absent.

We approach these open questions of dark atom cosmology in the present
work. Section \ref{sph} provides a brief review of papers on the properties of sphaleron(s) in models with heavy particle. The minimal Walking Technicolor (WTC) model is considered as an example. In section \ref{DAF} the dark atom formation at nucleosynthesis stage is discussed. The results are summarized in the Conclusion.

\section{Sphaleron in WTC model\label{sph}}

The static unstable solution of electroweak field equations in pure gauge theory was originally found by Manton in 1983 \cite{Manton} and called "sphaleron"\,. It corresponds to the saddle point at the top of the potential barrier separating topologically nonequivalent vacuums in configuration space. Author pointed that sphaleron arises as a consequence of the topology of the $SU(2)$ group. The main properties of the similar solution in the SM were considered in several papers throughout the 90s \cite{KunzBiSph, KunzAngle, Diakonov, KunzFermion}. There were found that
\begin{enumerate}
    \item Only for non-physical values of the Higgs parameters, the existence of additional branches of sphaleron transitions are possible \cite{KunzBiSph}.
    \item Sphaleron can be described with high accuracy using the spherically symmetric ansatz \cite{KunzAngle}. Its physical energy should be $E_{Sph} \approx 9.1 \, \mbox{TeV}$.
\end{enumerate}

The further study has shown that both these statements can be violated by physics beyond the SM. For instance, the existing of heavy fermions (with mass in order of $1 \, \mbox{TeV}$) leads to a significant change in the sphaleron energy \cite{KunzFermion}. The additional solution of field equations may arise as a consequence of a deformation of the Higgs potential \cite{SpannowskyTamarit}. Such changes should affect baryosynthesis due to sphaleron freezing out temperature $T_*$ decreases (or increases) \cite{OnAnomalous}. Actually the predicted ratio of DM and baryonic matter densities depends on the ratio $\frac{m_i}{T_*}$, where $m$ is a mass of "$i$" type particle. It means that the resulting uncertainty in the temperature value is insignificant and can be compensated by tuning unknown masses of heavy species. However, it is also necessary to compare the shifted sphaleron freezing out temperature $T_*$ with the temperature of electroweak phase transition $T_c$. The predictions may differ significantly for cases $T_*>T_c$ and $T_*<T_c$ \cite{WTCtermo1,WTCtermo2,WTCtermo3,HarveyTurner}.

Therefore, the determination of the exact properties of sphaleron transitions is needed to consider models suggesting the existence of dark atoms. This is especially true for the minimal WTC model which assumes both the new heavy fermions and a modified Higgs potential \cite{WTC1, WTC2, WTC3}. Unfortunately, studying the features of this SM extension involves some technical difficulties. Indeed, among others the low energy effective WTC model predicts 19 new (pseudo)scalar fields, which should be introduced into the system of equations. The problem is not only in the number of additional degrees of freedom, but also in the fact that the effect of spinless fields on the sphaleron solution has not yet been studied. All the same can be said with respect to composite vector fields.

However, it can be expected that this approach will allow to find more precise upper limits on the masses of heavy particles. Indeed, the height of the sphaleron potential barrier significantly decreases if new fermions are too heavy. This contradicts the absence of sphaleron transitions in modern experiments \cite{IceCubeLHC}. The influence of additional branches of sphaleron transitions on cosmological evolution is less obvious and requires a particular case study.

\section{Dark atom formation\label{DAF}}

The second step of the dark atom formation is serial capturing of light primordial nuclei $N^i$ by negatively charged heavy particles $X^{-2n}$. The simplest possible scenario assumes $n=2$. In that case dark atom $X^{--}He^{++}$, which is usually denoted as $OHe$, can be considered as a Bohr-like bound state with energy
\begin{equation}
    E_{OHe}^{\text{Bohr}}=8\alpha^2m_{He}\approx1.6\,\mbox{MeV}.
\end{equation}
It could easily estimated that O-helium is formed almost immediately after the formation of helium during standard nucleosynthesis \cite{Soldatov2010}. It is happens at temperatures $T\sim 1-100 \, \mbox{keV}$. 

However, in general case $n>1$ it is necessary to consider the much more complicated scenario. Indeed, there are four types of processes are possible at nucleosynthesis stage
\begin{align}
    &N^1+N^2\rightarrow N^3 +\gamma /N^4,
    \label{Reaction1}\\
    &X+N\rightarrow XN+\gamma,
    \label{Reaction2}\\
    &XN^{1}+N^{2}\rightarrow XN^3+\gamma/N^4,
    \label{Reaction3}\\
    &XN^1+XN^2\rightarrow X_2N^3+\gamma/N^4. \label{Reaction4}
\end{align}
The first of them (\ref{Reaction1}) describes the standard nuclear reactions. There are eleven main processes that produce helium-4 and some heavier nuclei (isotopes of lithium and beryllium) \cite{GorbRub}.
Reactions (\ref{Reaction2}) are similar to the standard recombination. In general case they produce negatively charged bound states (dark ions) $(XN)^{-2n+q_N}$, which are burns in type (\ref{Reaction3}) reactions. Unfortunately, this may lead to overproduction of anomalous isotopes and/or primordial metals \cite{AkhmedovPospelov}. The last type processes (\ref{Reaction4}) describes the merging of two dark ions into the molecule-like bound state. 

A correct description of nucleosynthesis in the presence of multicharged particles requires to consider each step of the dark atom formation. First of all, it is necessary to find the temperature when the (\ref{Reaction2}) type reactions become possible. By analogy with hydrogen recombination for non-relativistic equilibrium concentrations
\begin{equation}
    n_i^{\text{now}}\left(\cfrac{T}{T_{\text{now}}}\right)^3 = g_i \left(\cfrac{m_iT^{\frac{3}{2}}}{2\pi}\right)e^{-\frac{m_i}{T}},
\end{equation}
the Saha equation \cite{GorbRub} gives
\begin{equation}
    T_{\text{rec}}=E_{X-N}\left(\ln\left(\cfrac{g_Xg_N}{g_{XN}}\left(\cfrac{m_NT^2_{\text{now}}}{2\pi\, E_{X-N}}\right)^{\frac{3}{2}}\cfrac{1}{n_N^{\text{now}}}\right)\right)^{-1},
    \label{Trec}
\end{equation}
where $g_i$ is the the number of spin degrees of freedom, $m_N$ is the mass of captured light nucleus, $n_N^{\text{now}}$ is its present concentration and $E_{X-N}$ is a binding energy for the system ”heavy core plus nuclear shell”\,. The last one depends on the inner structure of the dark ion.

\begin{figure}[h]
    \centering{\includegraphics[width = 1\linewidth]{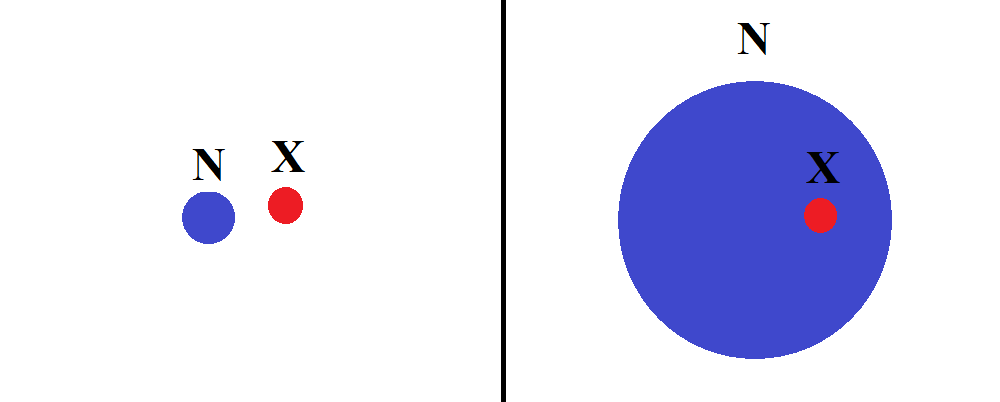}} 
    \caption{Bohr-like (left) and Thompson-like (right) dark atoms}
    \label{Fig:DAStructure}
\end{figure}

For high values of the charge parameter $n$ the ratio of nuclear and $X^{-2n}$ Bohr radii
\begin{equation}
    a=\cfrac{r_{N}}{r_B}\approx Z_X Z_N\, \alpha\, m_N\, r_0\, A_N^{1/3}
    \label{adef}
\end{equation}
may have values higher than one. This means that composite particle has not a Bohr, but Thomson structure (see Figure \ref{Fig:DAStructure}).
Table \ref{B-T} shows the results of calculations of some $XN$ bound states:
\begin{itemize}
    \item B --- all isotopes form Bohr-like particles;
    \item T --- all isotopes form Thomson-like particles;
    \item Number --- the mass number of the lightest Thomson-like particle.
\end{itemize}
There was accepted $r_0=1.3\cdot 5 \, \mbox{GeV}$. It can be noted that almost all neutral and positively charged states should have a Thomson structure, but hydrogen mostly forms Bohr-type dark ions. This is consistent with the result of solving the two-body Coulomb problem obtained in~\cite{AkhmedovPospelov}. The predicted structure of transitional states, which are indicated in the table by numbers, depends on value of $r_0$. For instance,  $OHe$ is Bohr-like dark atom if the values $r_0\sim(1.1-1.2)\cdot 5 \, \mbox{GeV}$ are accepted.
\begin{table}[h]
    \centering
    \begin{tabular}{|c|c|c|c|c|}
        \hline
            \multirow{2}{*}{$n$} & \multicolumn{4}{c|}{A} \\
        \hhline{~|----|}
            & $H$ & $He$ & $Li$ & $Be$ \\
        \hline
            1 & B & 4 & 3 & T \\
        \hline
            2 & 4 & 3 & T & T  \\
        \hline
            3 & 3 & T & T & T  \\
        \hline
            4 & 3 & T & T & T  \\
        \hline
            5 & 2 & T & T & T  \\
        \hline
    \end{tabular}
    \caption{The structure of the dark ions $XN$.}
    \label{B-T}
\end{table}

Therefore the Bohr energy 
\begin{equation}
    E_{X-N}^{\text{Bohr}}=2n^2Z^2\alpha^2m_N=\cfrac{1}{2m_N r_B^2}.
    \label{BohrBound}
\end{equation}
can be used in equation (\ref{Trec}) only for hydrogen and helium (in case $n=1$). To find the binding energy of other bound states, the harmonic oscillator approximation is often used in literature \cite{Soldatov2010, Glashow, KohriTakayama}. The Hamiltonian 
\begin{equation}
    \mathcal{H}=\begin{cases}
            \cfrac{p^2}{2m_N}-\cfrac{Z_X Z_N \alpha}{2r_N}\left(3-\cfrac{r^2}{r_N^2}\right), &r<r_N
            \\
            \cfrac{p^2}{2m_N}-\cfrac{Z_X Z_N \alpha}{r}, &r>r_N
            \end{cases}
    \label{GlashowHamiltonian}
\end{equation}
leads to the Thompson energy formula
\begin{equation}
    \begin{split}
         E_{X-N}^{\text{Thompson}}&=\cfrac{3}{2}\cfrac{Z_X Z_N \alpha}{r_N}\left(1-\sqrt{\cfrac{1}{Z_X Z_N \alpha m_N r_N}}\right)=\\
        &=\cfrac{3}{2}\cfrac{Z_X Z_N \alpha}{r_N}\left(1-\sqrt{\cfrac{r_B}{r_N}}\right),
    \end{split}
    \label{GlashowBound}
\end{equation}
which should be correct at region $2<a<\infty$. For intermediate values ($1<a<2$) it is proposed to find the binding energy variationally. The binding energy value of Bohr-like states is higher than for Thomson-like dark ions. This can be qualitatively explained by the compensation of Coulomb repulsion because of the spherical symmetry of the Hamiltonian (\ref{GlashowHamiltonian}).

Table \ref{tab:recomb} shows approximate values of dark ions recombination temperatures (start of reactions (\ref{Reaction2})) calculated with equations (\ref{BohrBound}) and (\ref{GlashowBound}). For the transitional cases ($D$ and $^4He$), only the boundary values are presented. The inner structure of the bound states is specified by the corresponding letter in brackets. The result depends weakly on the values of $g_X$ and $g_{XN}$. Since helium-4 forms at temperature $T\approx65\,\mbox{keV}$~\cite{GorbRub}, there are two possible scenarios are predicted:
\begin{itemize}
    \item For the $n<4$, the helium capturing happens earlier than protons and/or deuterium can be captured. Then, the excess of dark ions, $(XHe)^{-2n+2}$, is formed to start reactions (\ref{Reaction3}).
    \item For $n\geq4$, hydrogen capturing becomes possible before the $^4He$ formation. Therefore, another branch of reactions ($XH+N\rightarrow...$) is started. Such processes lead to an additional danger of  anomalous isotopes overproduction at later stages of the dark atom formation. This may help to set limits on the maximum charge of a heavy core.
\end{itemize}
Unfortunately, the temperature $T_{\text{rec}}$ for deuterium in case $n=5$ can not be estimated in this simple approach. Moreover, the drop in the energy value caused by the change of the inner structure can lead to non-trivial consequences.
\begin{table}[h]
    \centering
    \begin{tabular}{|c|c|c|c|c|}
        \hline
            \multirow{2}{*}{$n$} & \multicolumn{4}{c|}{$T_{\text{rec}},\,\mbox{keV}$} \\
        \hhline{|~|----|}
            & $p$ & $D$ & $^3He$& $^4He$\\
        \hline
            1& 3 (B) & 4 (B) & 28 (B) & 37 (B) / 6 (T) \\
        \hline
            2& 13 (B) & 19 (B) & $\sim$ 42 (T) & 85 (T) \\
        \hline
            3& 29 (B) & 44 (B) & $\sim$ 116 (T) &  180 (T) \\
        \hline
            4& 54 (B) & 79 (B) & $\sim$ 198 (T) &  285 (T) \\
        \hline
            5& 86 (B) & $\sim$ 126 (B) / 17 (T) & $\sim$ 286 (T) &  395 (T) \\
        \hline
    \end{tabular}
    \caption{Recombination temperatures of the dark atoms}
    \label{tab:recomb}
\end{table}

Since the standard reactions of nucleosynthesis and formation of dark atoms occur at the same temperatures, the Saha equation becomes inapplicable. Indeed, it requires that concentrations change only due to the expansion of the Universe, which is not true in considered case. Therefore, to describe processes \eqref{Reaction2}-\eqref{Reaction4} it is necessary to use the system of kinetic equations \cite{KhlopovBook}
\begin{equation}
    \cfrac{d n_i}{dt}+3 H n_i= \sum_{j,k} n_j n_k \left<\sigma v\right>^{jk}_i-n_i\sum_j n_j\left<\sigma v\right>^{ij},
    \label{system}
\end{equation}
where the energy-averaged cross sections $\left<\sigma v\right>^{jk}_i$ corresponds to $2\rightarrow2$ reactions $j+k\rightarrow i+l$. Also it should be mentioned that produced in such processes ordinary nuclei are not thermalized and should be described by additional equations
\begin{equation}
      \cfrac{\partial \phi_{N_i}}{\partial t}=\sum_{j,k} n_j n_k \cfrac{d (\sigma v)^{jk}_{N_i}}{dp_{N_i}}-\phi_{N_i} \sum_j n_j (\sigma v )^{{N_i}j}(p_{N_i})  
      -\phi_{N_i}\sum_j\int\phi_{N_j}(\sigma v)^{N_i N_j}dp_{N_j},
\end{equation}
where it is defined
\begin{equation}
    \cfrac{d n_N}{dp_N}=\phi_N(p_N,t).
\end{equation}

Finally, reactions \eqref{Reaction3} and \eqref{Reaction4} may freeze out before the nucleosynthesis stops. They are possible only if 
\begin{equation}
    n_{XN} \left<\sigma v\right> t >1.
    \label{MolFormCondMain}
\end{equation}
For a simple estimation, it can be assumed that
\begin{equation}
    n_{XN}\approx\cfrac{\rho_c}{m_{XHe}}\Omega_{DM}^{\text{now}}\left(\cfrac{T}{T_{\text{now}}}\right)^3,
\end{equation}
where $\Omega_{DM}^{\text{now}}$ is the observable energy density of dark matter. Also it  should be taken into account that radiation dominates. Therefore 
\begin{equation}
    t=\cfrac{3}{4}\cfrac{m_{Pl} }{T^2}\sqrt{\cfrac{5}{\pi^3 g_*}}.
\end{equation}
Here $m_{Pl}$ is the Plank mass and $g_*$ is a number of ultrarelativistic degrees of freedom at the considered temperature \cite{GorbRub}. For the non-relativistic particles "$i$" with the average thermal speed $\left<v\right>=\sqrt{\frac{8T}{\pi m_{i}}}$ the condition \eqref{MolFormCondMain} can be rewritten as (energy is measured in \mbox{GeV})
\begin{equation}
\begin{split}
    n_{XN} \left<\sigma v\right> t &\approx \cfrac{3}{2\pi^2}\cfrac{\rho_c \, \Omega_{DM}^{\text{now}} \, m_{Pl}\, \sigma}{m_{XHe} T_{\text{now}}^3}\sqrt{\cfrac{10\, T^3}{g_* \, m_{i}}}\approx 
    \\
    &\approx 3\cdot 10^9 \, \cfrac{\sigma}{m_{XN}\sqrt{m_{i}}} \, T^{\frac{3}{2}}>1.
\end{split}
    \label{condition}
\end{equation}

If the cross section $\sigma$ has a typical nuclear value
\begin{equation}
    \sigma \sim \pi r_{XN}^2 \approx 10^{-25}\,\mbox{cm}^2 = 250\, \mbox{GeV}^{-2},
    \label{sigOHe}
\end{equation}
then equation \eqref{condition} predicts the similar scenario for both types of reactions. This condition is fulfilled at high temperatures $T \sim 50-100\, \mbox{keV}$, but strongly violated at $T \sim 1 \, \mbox{keV}$. The rate difference between the reactions of \eqref{Reaction3} and \eqref{Reaction4} types is determined by the ratio of masses ${m_{XHe}}/{m_{\text{N}}}$. The production of dark molecules should stop earlier then capturing of light nuclei. Moreover, if masses of heavy particles is high enough ($m_X\sim 10\,\mbox{TeV}$), the formation of molecule-like states $X_2N$ does not start.

The main source of inaccuracy in considered estimation is the value of cross section \eqref{sigOHe}. Its exact value is unknown for the most reactions therefore it is impossible to make a correct prediction of the modified nucleosynthesis result.

\section{Conclusion}

The dark atoms $XHe$ are the composite DM candidates. The cosmological consequences of this model are studied pretty well for the simplest case when the heavy particles $X$ are only doubly charged. However, the description of dark atom formation is still incomplete. On the one hand, the predicted density of DM significantly depends on the properties of sphaleron transitions, which may change in different extensions of the SM. On the other hand, the process of light nuclei serial capturing has not been sufficiently studied.

To solve these two problems of considered model it is necessary to
\begin{itemize}
    \item find the sphaleron solution of the field equations for the particular extension of the SM and then to estimate the freezing out temperature of  sphaleron transitions;
    \item find the solution of the system of kinetic equations which describes the modified nucleosynthesis.
\end{itemize}
The computational complexity of such calculations is complemented by the lack of an accurate quantum description of the Thompson-like dark atoms.

\section*{Acknowledgements}
The research by M.Y.K. and D.O.S. was carried out in Southern Federal University with financial support from the Ministry of Science and Higher Education of the Russian Federation (State contract GZ0110/23-10-IF).

%% The bibliography section

\end{document}